\documentclass[12pt]{article}
\pdfoutput=1
\usepackage[english]{babel}
\usepackage[normalem]{ulem}
\usepackage{multirow}
\usepackage{booktabs}

\usepackage[text={17.0cm,23.7cm}]{geometry}
\usepackage{soul}
\usepackage{graphicx}
\usepackage[dvipsnames]{xcolor}
\usepackage{pdfcolmk}
\usepackage{amsmath}
\usepackage{amssymb}

\usepackage{eurosym}
\usepackage{slashed}
\usepackage{hyperref}
\usepackage{wasysym}
\usepackage{float}
\usepackage{placeins}
\usepackage{rotating}
\usepackage{tabularx}
\usepackage{cite}

\newcommand{\DM}{\mathrm{DM}}

\newcommand{\GeV}{~\mathrm{GeV}}
\newcommand{\TeV}{~\mathrm{TeV}}
\newcommand{\cm}{~\mathrm{cm}}

\newcommand{\sdp}{\ensuremath{\sigma_{\text{SD}}^{p}\,}}

\newcommand{\sip}{\ensuremath{\sigma_{\text{SI}}^{p}\,}}

\newcommand{\fv}{\ensuremath{f(\vec{v})}}

\begin{document}

\title{\vspace{-2cm}
{\normalsize
\flushright TUM-HEP 1000/15\\}
\vspace{0.6cm}
\bf A novel approach to derive halo-independent limits on dark matter properties \\ [8mm]}

\author{Francesc Ferrer$^1$, Alejandro Ibarra$^2$, Sebastian Wild$^2$\\[2mm]
{\normalsize\it $^1$ Physics Department and McDonnell Center for the Space Sciences,}\\[-0.05cm]
{\it\normalsize Washington University, St Louis, MO 63130, U.S.A.}\\[2mm]
{\normalsize\it $^2$ Physik-Department T30d, Technische Universit\"at M\"unchen,}\\[-0.05cm]
{\it\normalsize James-Franck-Stra\ss{}e, 85748 Garching, Germany}
}

\maketitle
\begin{abstract}
We propose a method that allows to place an upper limit on the dark matter elastic scattering cross section with nucleons which is independent of the velocity distribution. Our approach combines null results from direct detection experiments with indirect searches at neutrino telescopes, and goes beyond previous attempts to remove astrophysical uncertainties in that it directly constrains the particle physics properties of the dark matter. The resulting halo-independent upper limits on the scattering cross section of dark matter are remarkably strong and reach $\sigma_{\text{SI}}^{p} \lesssim 10^{-43} \, (10^{-42})~\text{cm}^2$ and  $\sigma_{\text{SD}}^{p} \lesssim 10^{-37} \, (3\times 10^{-37 })~\text{cm}^2$, for dark matter particles of $m_{\text{DM}}\sim 1~\text{TeV}$ annihilating into $W^+W^-$ ($b\bar b$), assuming $\rho_\text{loc}=0.3\text{ GeV}/\text{cm}^3$.
\end{abstract}

\section{Introduction}
\label{sec:intro}

Save for its gravitational effects, which establish dark matter (DM) as the 
second largest contributor to the energy budget of the Universe~\cite{Ade:2015xua}, very little 
is known about its composition~\cite{Bertone:2010zza,Bergstrom:2000pn,Jungman:1995df,Bertone:2004pz}. An attractive candidate for
making up the dark matter is a weakly interacting massive particle (WIMP),
that could be unveiled through its weak-scale interactions. 

Two complementary approaches have been proposed to probe the hypothetical WIMP population inside the Solar System. Direct detection experiments aim to detect the nuclear recoil induced by the elastic scattering of the dark matter particles traversing a detector at the Earth~\cite{Goodman:1984dc}.
The second approach consists in the search using neutrino telescopes of the high energy neutrinos which are hypothetically produced in the annihilation of dark matter particles which have been previously captured in the Sun via a series of scatterings with the solar matter~\cite{Silk:1985ax}. 

Over the past few years, these programs have attained the sensitivity to probe
the particle physics properties in a vast range of WIMP scenarios. 
The interpretation of these experimental efforts, however, is hampered 
by uncertainties in the required astrophysical input. For instance, 
the event rates at both neutrino telescopes (NT) and direct detection (DD) 
experiments are sensitive 
to the elastic DM-nucleon scattering cross section. But 
to translate experimental data into upper limits on the DM scattering
cross section requires the input of a DM velocity distribution \fv. 

A common strategy is to adopt a Maxwell-Boltzmann velocity distribution in 
the galactic rest frame.
Using this scheme, upper limits on the spin-independent and spin-dependent WIMP-nucleon scattering cross section have been published by direct detection experiments, such as  LUX~\cite{Akerib:2013tjd}, XENON100~\cite{Aprile:2012nq}, SuperCDMS~\cite{Agnese:2014aze}, PICO~\cite{Amole:2015lsj}, COUPP~\cite{Behnke:2012ys} and SIMPLE~\cite{Felizardo:2011uw}, as well as by neutrino telescopes, such as Super-Kamiokande~\cite{Choi:2015ara} and IceCube~\cite{Aartsen:2012kia}. 
A Maxwell-Boltzmann velocity distribution, however, only holds when the dark matter distribution is modeled after the Standard Halo Model (SHM)~\cite{Drukier:1986tm}, an isothermal sphere profile with an isotropic velocity distribution. Recent N-body simulations, however, provide evidence that a Maxwellian distribution is not a good description of the smooth halo component~\cite{Kuhlen:2009vh,Lisanti:2010qx,Mao:2012hf}. Furthermore, the dark matter halo of our Galaxy might contain tidal streams or a dark disk component ~\cite{Read:2008fh,Read:2009iv,Purcell:2009yp,Ling:2009eh}, adding a new source of uncertainty to the dark matter velocity distribution inside the Solar System. 
Therefore, to robustly constrain the WIMP scattering cross section from experiments, it is necessary to devise methods that do not depend on the choice of the velocity distribution.

Several methods have been proposed to derive halo-independent statements from direct detection experiments~\cite{Fox:2010bu,Fox:2010bz,McCabe:2010zh,McCabe:2011sr,Frandsen:2011gi,Gondolo:2012rs,HerreroGarcia:2012fu,DelNobile:2013cva,Fox:2014kua,Feldstein:2014gza,Feldstein:2014ufa,Anderson:2015xaa,Bozorgnia:2014gsa}.
Broadly speaking, the rationale behind these studies can be divided into two categories. First, the experimental data can be expressed in terms of a variable that includes both the information about the DM scattering cross section and the (integrated) velocity distribution, often denoted as $\tilde{\eta}(v_\text{min})$. By comparing measurements and upper limits of such an observable from different direct detection experiments, one can quantify the compatibility of a positive claim with a null result in a halo-independent way. Second, the existing methods can be used to infer fundamental particle physics properties of DM from positive signals in current or future experiments. Along these lines, it has also been shown that the complementarity of direct searches and neutrino telescopes can be important, due to the different dependence of both detection methods on the DM velocity distribution~\cite{Kavanagh:2014rya,Blennow:2015oea}.

In this work we propose a novel method which allows for the first time to use existing data for calculating an \emph{upper limit} on the scattering cross section which is independent of the velocity distribution, by combining null results from direct detection experiments and neutrino telescopes. 

The paper is organized as follows. In section~\ref{sec:formalism} we outline our method and in section~\ref{sec:superposition} we derive an upper limit on the elastic scattering cross section from a direct detection experiment or from a neutrino telescope for a general velocity distribution. Then, in section~\ref{sec:limits_arbitrary_f} we apply our method to derive an upper bound on the cross section which is independent of the velocity distribution. This is the main result of this paper. Lastly, in section~\ref{sec:lower_limit_xenon1t}, we apply our method to derive a halo-independent lower limit on the scattering cross section from a hypothetical signal in a future direct detection experiment. We present our conclusions in section~\ref{sec:conclusions}, and provide details of direct detection data in appendix~\ref{sec:appendix_dd}.

\section{Probing WIMP dark matter inside the Solar System} 
\label{sec:formalism}

We assume that the Solar system is embedded in a distribution of WIMP dark matter particles with mass density $\rho_\text{loc}$. The DM halo is spatially homogeneous on Solar system scales and its velocity distribution \fv\  {\em in the rest frame of the Sun} is normalized as
\begin{align}
\int_{v\leq v_\text{max}}  \text{d}^3 v \fv =1\;,
\label{eq:normalization}
\end{align}
where $v\equiv|\vec v|$ and $v_\text{max}$ is the maximal velocity of the dark matter particles in the galactic halo. 
Under the common assumption that all dark matter particles in the halo are gravitationally bound to the Galaxy, $v_\text{max} \simeq 777$ km/s, which is the sum of the galactic escape velocity $\simeq 533$ km/s~\cite{Piffl:2013mla} and the local velocity of the Sun with respect to the halo $\simeq 244$ km/s~\cite{Xue:2008se,McMillan:2009yr,Bovy:2009dr}. We comment in section~\ref{sec:limits_arbitrary_f} on the implications of relaxing this assumption.

The number of expected recoil events at a DD experiment can be expressed as:
\begin{align}
	R=\mathcal{E} \cdot \sum_i \int_{0}^\infty \text{d}E_R \, \epsilon (E_R) \frac{\xi_i \rho_\text{loc}}{m_{A_i} m_{\text{DM}}} \int_{v \geq v_{\text{min},i}^{(\text{DD})}(E_R)} \text{d}^3 v \, v f (\vec{v}+\vec{v}_{\rm obs}(t)) \, \frac{\text{d}\sigma_i}{\text{d}E_R} \,.
\label{eq:diff_scattering_rate}
\end{align}
Here, $\vec v$ denotes the dark matter velocity in the detector frame, hence the velocity distribution of dark matter particles is $f (\vec{v}+\vec{v}_{\rm obs}(t))$, with $\vec{v}_{\rm obs}(t) \equiv \vec{v}_\oplus(t) $ being the velocity of the Earth with respect to the solar frame, and $\left| \vec{v}_\oplus(t) \right| = 29.8$ km/s (we note again that we define $f(\vec{v})$ in the rest frame of the Sun). Furthermore, $\text{d}\sigma_i/\text{d}E_R$ is the differential scattering cross section of a WIMP off a nuclear isotope $i$ with mass $m_{A_i}$ and mass fraction $\xi_i$ in the detector, and $v_{\text{min},i}^{(\text{DD})}(E_R) = \sqrt{m_{A_i} E_R/(2 \mu_{A_i}^2)}$ is the minimal speed necessary for a dark matter particle to induce a recoil with energy $E_R$,  with $\mu_{A_i}$ being the reduced mass of the WIMP-nucleus scattering. Lastly, $\epsilon(E_R)$ and $\mathcal{E}$ are the detection efficiency and exposure, respectively, both depending on the specific experiment under consideration. 

On the other hand, the neutrino flux from annihilations in the Sun is completely determined by the capture rate, assuming that dark matter capture and annihilation are in equilibrium. The capture rate  is given by~\cite{Gould:1987ir}
\begin{align}
	C=\sum_i \int_0^{R_\odot} 4\pi r^2 \text{d}r \, \eta_i(r) \frac{\rho_\text{loc}}{m_\text{DM}} \int_{v \leq v_{\text{max},i}^{\text{(Sun)}}(r)} \text{d}^3 v \, \frac{ f (\vec{v})}{v} &\left(v^2+\left[v_\text{esc}(r)\right]^2 \right) \nonumber \times \\ 
	&\int_{m_\text{DM} v^2 /2}^{2 \mu_{A_i}^2 \left(v^2+\left[v_\text{esc}(r)\right]^2 \right)/m_{A_i}} \text{d} E_R \, \frac{\text{d} \sigma_i}{\text{d}E_R} \,,
\label{eq:general_formula_capture_rate}
\end{align}
where $\eta_i(r)$ is the number density of the element $i$ at a distance $r$ from the solar center, $v_\text{esc}(r)$ is the escape velocity, and $v_{\text{max},i}^{\text{(Sun)}}(r) = 2 \, v_\text{esc}(r) \sqrt{m_\text{DM} m_{A_i}}/\left| m_\text{DM} - m_{A_i}\right|$  is the maximum speed of a dark matter particle such that the capture in the Sun remains kinematically possible. For scattering cross sections that can be currently probed by neutrino telescopes, equilibration is achieved as long as the annihilation cross section multiplied by the relative dark matter velocity satisfies $( \sigma v )_{\text{ann}} \gtrsim 10^{-28} \text{ cm}^3/\text{s}$ \footnote{Capture and annihilation in the Sun are in equilibrium if $\alpha_\text{eq} \equiv \tanh^2 \left( t_\odot \sqrt{C \cdot \Gamma_A} \right)$ is close to one, where $t_\odot \simeq 1.5 \cdot 10^{17}$~s is the age of the Sun, and $\Gamma_A \simeq 5.17 \cdot 10^{-57} \frac{1}{\text{s}} \cdot \frac{\langle \sigma v \rangle}{3 \cdot 10^{-26} \text{cm}^3/\text{s}} \cdot \left( \frac{m_\text{DM}}{\text{GeV}} \right)^{3/2}$ is the annihilation constant~\cite{Jungman:1995df}. It is straightforward to check that for $( \sigma v )_{\text{ann}} \gtrsim 10^{-28} \text{ cm}^3/\text{s}$ and for the capture rates $C$ that are currently probed by IceCube or Super-Kamiokande, one has $\alpha_\text{eq} \gtrsim 0.99$.}. In the rest of this work, we will always assume that equilibrium has been reached.

The largest uncertainties in the calculation of the scattering and the capture rates in a given particle physics model stem from our ignorance of the WIMP scattering cross section with nuclei and of the dark matter velocity distribution. It is common in the literature to cast the differential cross section as~\cite{Cerdeno:2010jj}
\begin{align}
\frac{\text{d}\sigma_i}{\text{d}E_R}=\frac{m_{A_i}}{2\mu_{A_i}^2 v^2}
(\sigma_\text{SI}F_{i,\text{SI}}^2(E_R)+\sigma_\text{SD}F_{i,\text{SD}}^2(E_R)) \,,
\label{eq:x-section}
\end{align}
where $\sigma_\text{SI}$ and $\sigma_\text{SD}$ are, respectively, the spin-independent and spin-dependent cross sections at zero momentum transfer, which can be calculated in a concrete dark matter model in terms of its fundamental parameters, while $F_{i,\text{SI}}(E_R)$ and $F_{i,\text{SD}}(E_R)$  are form factors that depend on the nucleus, and which are given in~\cite{Lewin:1995rx} and in~\cite{Bednyakov:2006ux} for the spin-independent and the spin-dependent interactions, respectively. Furthermore, most analyses adopt a Maxwell-Boltzmann velocity distribution in the galactic rest frame. Under this assumption, the null searches for dark matter particles in the Solar System using direct detection experiments and from neutrino telescopes are translated into limits on $\sigma_\text{SI}$ and $\sigma_\text{SD}$, and confronted with the predictions from models. Clearly, a particle physics model violating these limits is not necessarily excluded, since the actual velocity distribution might deviate from the simple Maxwellian form. It is then important to derive limits on $\sigma_\text{SI}$ and $\sigma_\text{SD}$ which do not depend on the form of the velocity distribution, in order to robustly exclude a concrete particle physics scenario. In the following sections we will present a novel method to obtain halo-independent limits on the cross section which exploits the complementarity between direct detection experiments and neutrino telescopes.

\section{Dark matter in the solar system as a superposition of streams}
\label{sec:superposition}

For the purpose of deriving limits on the WIMP cross section for general velocity distributions we find it convenient to decompose the dark matter population in the solar system as
\begin{align}
f(\vec v)= \int_{|\vec v_0|\leq v_\text{max}} \text{d}^3 v_0 \,\delta^{(3)}(\vec v-\vec v_0) f(\vec v_0) \;,
\label{eq:f_decomp}
\end{align}
which physically can be interpreted as a superposition of hypothetical streams with fixed velocity $\vec v_0$ with respect to the solar frame, $f_{\vec v_0}(\vec v)=\delta^{(3)}(\vec v-\vec v_0)$. Then, using Eqs.~(\ref{eq:diff_scattering_rate}) and (\ref{eq:general_formula_capture_rate}), the number of expected scattering events $R$ and the capture rate $C$ can be cast as
\begin{align}
R &=\int_{|\vec v_0|\leq v_\text{max}} \text{d}^3 v_0 \, f(\vec v_0) \, R_{\vec v_0}\;, \nonumber \\
C&=\int_{|\vec v_0|\leq v_\text{max}} \text{d}^3 v_0\, f(\vec v_0) \, C_{\vec v_0} \;,
\label{eq:R-C-decomp}
\end{align}
where $R_{\vec v_0}$ and $C_{\vec v_0}$ are, respectively, the number of scattering events and the capture rate for the dark matter stream with velocity distribution $f_{\vec v_0}(\vec{v})=\delta^{(3)}(\vec v -\vec v_0)$. 

\begin{figure}
\begin{center}
\hspace{-0.5cm}
\includegraphics[scale=1.05]{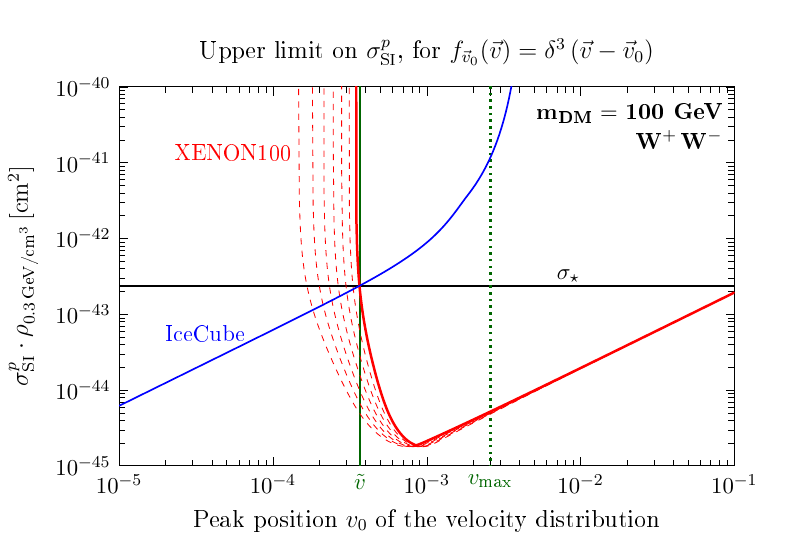}\hspace{0.2cm} 
\includegraphics[scale=1.05]{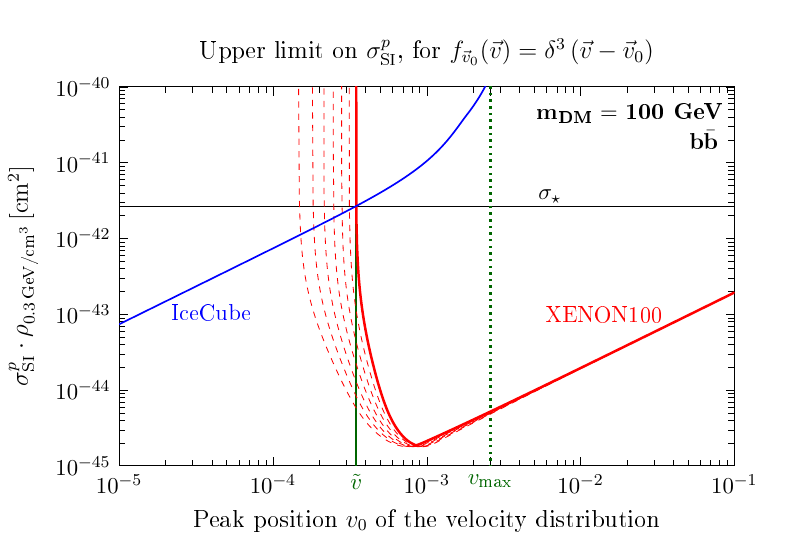}\\
\hspace{-0.5cm}
\includegraphics[scale=1.05]{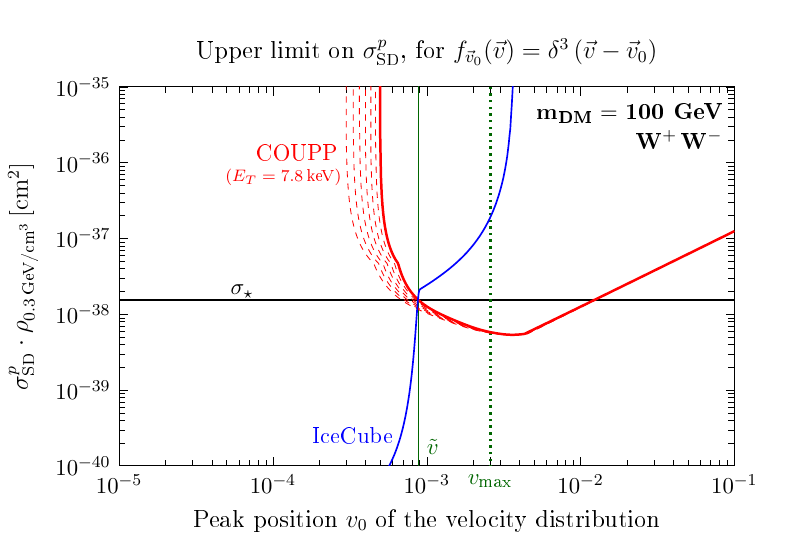}\hspace{0.2cm} 
\includegraphics[scale=1.05]{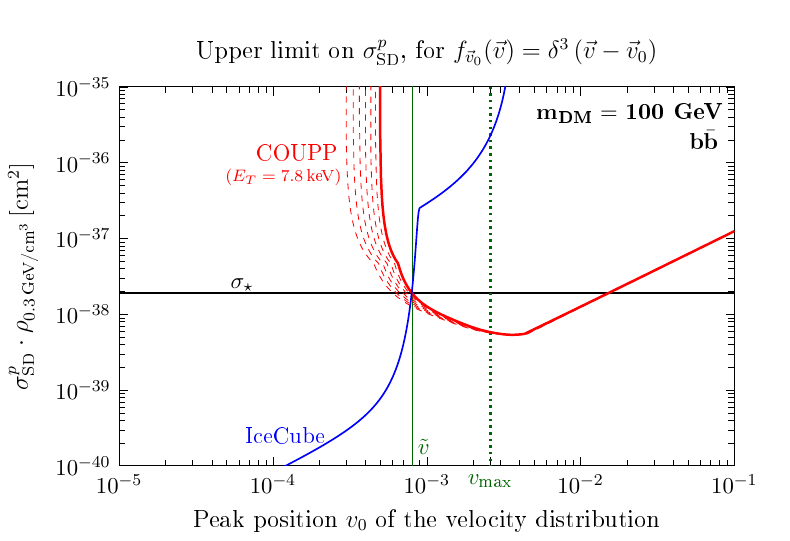}
\end{center}
\caption{\small Upper limits on the spin-independent dark matter-proton scattering cross section $\sip$, normalized to a local dark matter density $\rho_\text{loc} = 0.3 \text{ GeV}/\text{cm}^3$, derived from the null results of XENON100 and IceCube  (upper plots)  and on the spin-dependent dark matter-proton scattering cross section $\sdp$ from the null results of COUPP and IceCube (lower plots), assuming annihilations into $W^+W^-$ (left plots) or $b\bar b$ (right plots), for $m_\text{DM}=100$ GeV and for a velocity distribution corresponding to a stream with speed $v_0$ with respect to the rest frame of the Sun, $f_{\vec v_0}(\vec{v})=\delta^{(3)}(\vec v-\vec v_0)$. The various dashed red lines denote the upper limits for different angles between $\vec{v}_0$ and the velocity of the Earth. In each plot, the maximal value of the cross section allowed by the direct detection experiment and the neutrino telescope is denoted by $\sigma_*$. 
}
\label{fig:limits-streams}
\end{figure}

For every stream with velocity $\vec v_0$ with respect to the Sun (and therefore with velocity $\vec v_0 -\vec v_E(t)$ with respect to the Earth) one can obtain an upper limit on the scattering cross section from the null results of a direct detection experiment by requiring that $R_{\vec v_0}\leq R_\text{max}$. Here, $R_\text{max}$ is the upper limit on the expected number of scattering events following from the results of the corresponding experiment. For definiteness, we assume equal coupling to protons and neutrons for the spin-independent scattering and coupling only to protons for the spin-dependent scattering. Appendix~\ref{sec:appendix_dd} provides further details of the calculation. We show in Fig.~\ref{fig:limits-streams}, for illustration, the  90\% C.L. upper limits on the spin-independent (upper panels) and spin-dependent  cross sections (lower panels) for $m_\text{DM}=100\GeV$ as a function of the stream speed $v_0\equiv |\vec v_0|$ which follow from the null results from XENON100 or COUPP, respectively. 
The dashed red lines show the limits for various angles between the stream velocity $\vec v_0$ and a fixed Earth velocity $\vec v_E$, while the red solid line corresponds, for a fixed speed $v_0$, to the weakest among the limits for all angles. This conservative upper limit on the cross section will be denoted as $\sigma^\text{DD}_\text{max}(v_0)$, either for the spin-independent or the spin-dependent cross section. Then, by construction,  $R_{\vec v_0}(\sigma)\geq R_\text{max}$ for $\sigma\geq \sigma^\text{DD}_\text{max}(v_0)$.

Similarly, for every stream there is an upper limit on the scattering cross section which is allowed by a neutrino telescope, which we calculate from the requirement that $C_{\vec v_0}\leq C_\text{max}$, where $C_\text{max}$ is the maximum capture rate allowed for a given annihilation channel, assuming equilibration between capture and annihilation. For the calculation of the capture rate induced by the spin-independent coupling we include the 29 most abundant elements in the Sun and we assume Gaussian form factors, as parametrized in~\cite{Gondolo:2004sc}. 
On the other hand, to obtain the capture rate induced by the spin-dependent coupling to the proton we take into account, following~\cite{Catena:2015uha}, not only the scattering off hydrogen, as commonly done in the literature, but also off $^{14}$N, with form factors as given in~\cite{Catena:2015uha}.  
For the number density profile we adopt the solar model AGSS09~\cite{Serenelli:2009yc}. The blue solid line in Fig.~\ref{fig:limits-streams} shows the 90\% C.L. upper limits on the spin-independent (upper panels) and spin-dependent cross sections (lower panels) for dark matter streams with speed $v_0\equiv |\vec v_0|$ which follow from the requirement that the capture rate in the Sun is smaller than the maximum value $C_\text{max}$ inferred from the IceCube data. We take again $m_\text{DM}=100$ GeV, and consider the annihilation final states $W^+W^-$ (left panels) and $b\bar b$ (right panel). The noticeable kink in the IceCube limit on the spin-dependent cross section is due to the fact that at low velocities the capture rate is dominated by scatterings off H, while scatterings off $^{14}$N become important at large velocities. In the case of neutrino telescopes, the most conservative upper limit on the cross section for a given stream speed $v_0$ which follows from the null results of a neutrino telescope will be denoted as $\sigma^\text{NT}_\text{max}(v_0)$, therefore, $C_{\vec v_0}(\sigma)\geq C_\text{max}$ for $\sigma\geq \sigma^\text{NT}_\text{max}(v_0)$.

The limits on the scattering cross section from direct detection or from neutrino telescopes derived for streams with speed $v_0$ can be used to calculate limits for a general velocity distribution $f(\vec v)$. Noting that the differential cross section is linear in $\sigma_\text{SI}$, $\sigma_\text{SD}$, it follows that the number of expected scattering events satisfies
\begin{align}
R_{\vec v_0}(\sigma)=\frac{\sigma}{\sigma^\text{DD}_\text{max}(v_0)} R_{\vec v_0}[\sigma^\text{DD}_\text{max}(v_0)]\geq \frac{\sigma}{\sigma^\text{DD}_\text{max}(v_0)} R_\text{max}\;,
\end{align}
where $\sigma$ denotes either $\sigma_\text{SI}$ or $\sigma_\text{SD}$, and we have used the fact that, for the dark matter stream with velocity $\vec v_0$, $\sigma^\text{DD}_\text{max}(v_0)$ gives the largest scattering rate allowed by the direct detection experiment. Inserting this inequality in Eq.~(\ref{eq:R-C-decomp}) gives
\begin{align}
R(\sigma)\geq \int_{|\vec v_0|\leq v_\text{max}} \text{d}^3 v_0 f(\vec v_0) \frac{\sigma}{\sigma^\text{DD}_\text{max}(v_0)} R_\text{max}\;.
\end{align}
Finally, requiring that the number of expected events induced by the cross section $\sigma$ is in agreement with the experimental upper limit, $R(\sigma)\leq R_\text{max}$, one obtains from a given direct detection experiment the upper limit on the cross section 
\begin{align}
\sigma \leq \left[\int_{|\vec v_0|\leq v_\text{max}}  \text{d}^3 v_0 \frac{f(\vec v_0)}{\sigma^{\rm DD}_{\rm max}(v_0)}\right]^{-1}\;.
\label{eq:sigmaDD_general}
\end{align}
A similar rationale applied to the capture rate at a neutrino telescope gives 
\begin{align}
\sigma \leq \left[\int_{|\vec v_0|\leq v_\text{max}} \text{d}^3 v_0 \frac{f(\vec v_0)}{\sigma^{\rm NT}_{\rm max}(v_0)}\right]^{-1}
\label{eq:sigmaNT_general}\;.
\end{align}
This formalism allows to readily calculate upper limits on the cross section for a general velocity distribution from the function $\sigma_\text{max}(v_0)$ derived for a given direct detection experiment or neutrino telescope, without the necessity of going through the rather involved calculations described in Section \ref{sec:formalism}. \footnote{The functions $\sigma^\text{DD}_\text{max}(v_0)$ and $\sigma^\text{NT}_\text{max}(v_0)$ are available from the authors upon request.}

\section{A halo-independent upper limit on the scattering cross section}
\label{sec:limits_arbitrary_f}

Direct detection experiments and neutrino telescopes probe the WIMP population in the Solar System in a complementary way:  direct detection experiments are insensitive to slow moving WIMPs, since the recoiling nucleus has an energy below the detector threshold. However, this population of dark matter particles can be efficiently captured inside the Sun and thus leads to a high energy neutrino flux which could be detected in a neutrino telescope. This complementarity is apparent from Fig.~\ref{fig:limits-streams} for dark matter streams, since for {\it every} stream speed $v_0$ there exists a finite upper limit on the cross section, which follows either from a direct detection experiment or from a neutrino telescope.

We will show in this section that the complementarity between direct detection experiments and neutrino telescopes makes it possible to derive an upper limit on the scattering cross which is independent of the velocity distribution by using the upper limits on the cross section for dark matter streams. To this end, we first determine, for a given dark matter mass and a given annihilation channel, the largest value of the scattering cross section $\sigma_*$ which is allowed by a concrete direct detection experiment {\it and} a concrete neutrino telescope, assuming a stream-like velocity distribution with speed between 0 and $v_\text{max}$:
\begin{align}
\sigma_* \equiv 
\text{max}\left\{\sigma^\text{DD}_\text{max}(\tilde v), \sigma^\text{DD}_\text{max}(v_\text{max})\right\}\;,
\label{eq:sigma_star}
\end{align}
where $\tilde v $ is the speed for which  $\sigma^\text{DD}_\text{max}(\tilde v)=\sigma^\text{NT}_\text{max}(\tilde v)$. The determination of the maximal scattering cross section allowed by a direct detection experiment and by a neutrino telescopes is illustrated in Fig.~\ref{fig:limits-streams} for a dark matter mass $m_\text{DM}=100$ GeV which annihilates into $W^+W^-$ (left panels) and $b\bar b$ (right panels), considering XENON100 and IceCube for the spin-independent limits (upper panels), and IceCube and COUPP for the spin-dependent limits (lower panels). Then, by construction, 
\begin{align}
	\sigma_\text{max}^\text{DD}(v_0)\leq \sigma_* & \quad \text{for}~ \tilde{v} \leq v_0 \leq v_\text{max} \;,
\label{eq:bound_sigmav0_DD}\\
\sigma_\text{max}^\text{NT}(v_0)\leq \sigma_*  &  \quad \text{for}~ 0\leq v_0 \leq \tilde{v}
\label{eq:bound_sigmav0_NT}\;.
\end{align}

To derive an upper limit on the cross section by combining results from a direct detection experiment and a neutrino telescope we first note from Eq.~(\ref{eq:sigmaDD_general}) that
\begin{eqnarray}
\sigma &\leq& \left[\int_{0\leq v_0\leq v_{\rm max}} \text{d}^3 v_0 \frac{f(\vec v_0)}{\sigma^{\rm DD}_{\rm max}( v_0)}\right]^{-1}  \nonumber \\
&\leq& \left[\int_{\tilde v\leq v_0\leq v_{\rm max}} \text{d}^3 v_0 \frac{f(\vec v_0)}{\sigma^{\rm DD}_{\rm max}( v_0)}\right]^{-1} \;.
\end{eqnarray}
Using now Eq.~(\ref{eq:bound_sigmav0_DD}) and defining
\begin{align}
\delta_f \equiv \int_{\tilde v \leq v_0 \leq v_\text{max}} \text{d}^3 v_0 \, f( \vec{v}_0 )\;,
 \label{eq:deltaf_definition}
\end{align}
we obtain 
\begin{equation}
\sigma \leq 
\frac{\sigma_*}{\delta_f}\;.
\label{eq:bound_sigma_DD}
\end{equation}

An analogous calculation applied to the upper limit on the cross section from a neutrino telescope, Eq.~(\ref{eq:sigmaNT_general}), gives, using Eq.~(\ref{eq:bound_sigmav0_NT}),
\begin{equation}
\sigma \leq \frac{\sigma_*}{1-\delta_f}\;,
\label{eq:bound_sigma_NT}
\end{equation}
where the factor $(1-\delta_f)$ follows from the normalization of the velocity distribution, Eq.~(\ref{eq:normalization}), and the definition of $\delta_f$, Eq.~(\ref{eq:deltaf_definition}).

Finally, the upper limits on the cross section from a direct detection experiment and from a neutrino telescope,  Eqs.~(\ref{eq:bound_sigma_DD}) and (\ref{eq:bound_sigma_NT}), imply
\begin{align}
\sigma\leq 2\sigma_*\;.
\label{eq:main_result}
\end{align}
This is the main result of this paper: for a given dark matter mass and a given annihilation channel, there exists an upper limit on the scattering cross section which follows from combining the null results from direct detection experiments and neutrino telescopes and which is independent of the choice of the halo velocity distribution $f(\vec v)$. In this equation, $\sigma_*$ can be calculated following the procedure described in Section \ref{sec:formalism} from the analysis of dark matter streams. It can also be checked that the distributions that saturate the limit Eq.~(\ref{eq:main_result}) satisfy $\delta_f=1/2$.

\begin{figure}[t!]
\begin{center}
\hspace{-0.8cm}
\includegraphics[scale=0.95]{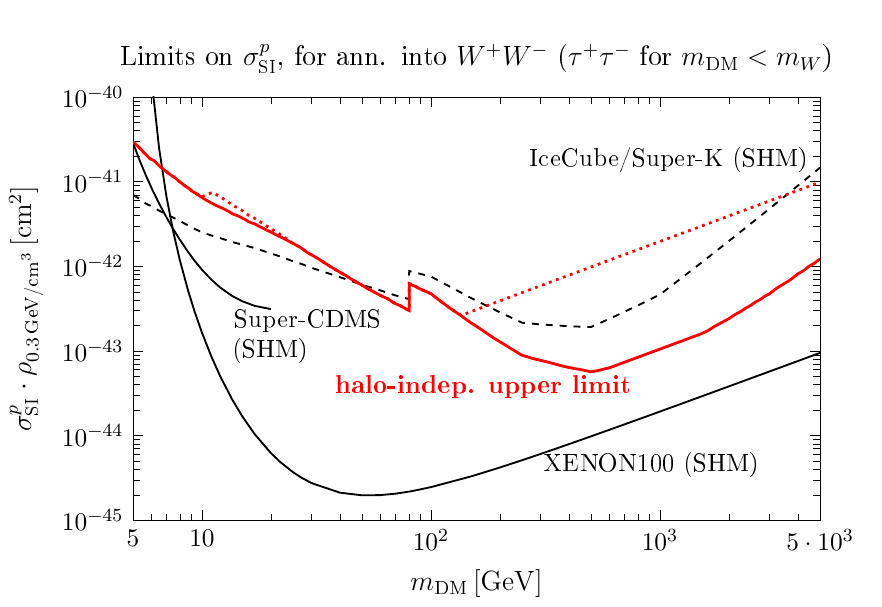}\hspace{0.05cm}
\includegraphics[scale=0.95]{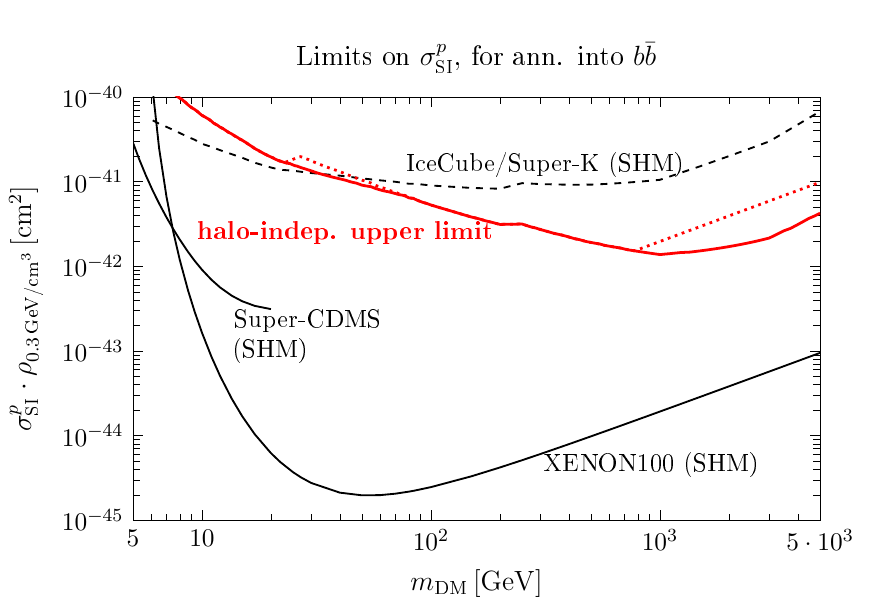}\\
\hspace{-0.8cm}
\includegraphics[scale=0.95]{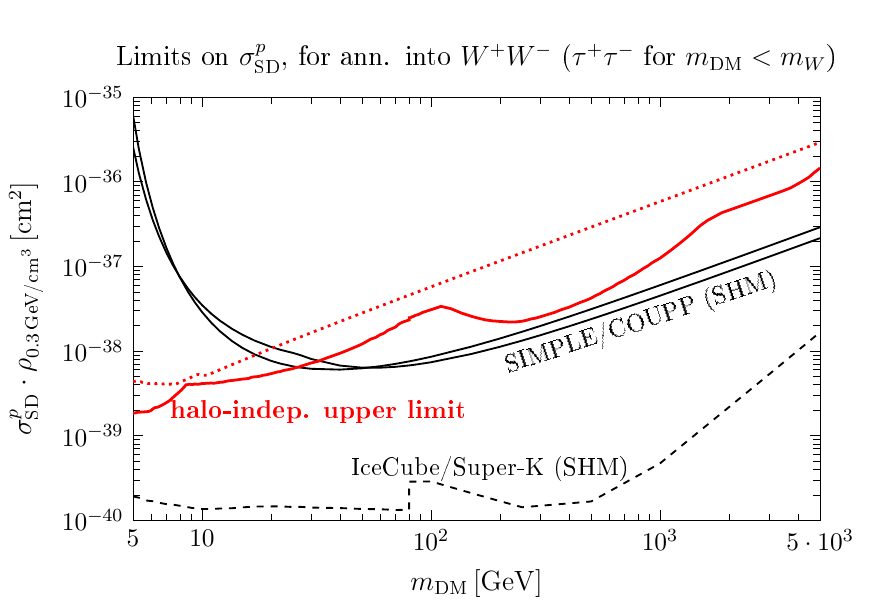}\hspace{0.05cm}
\includegraphics[scale=0.95]{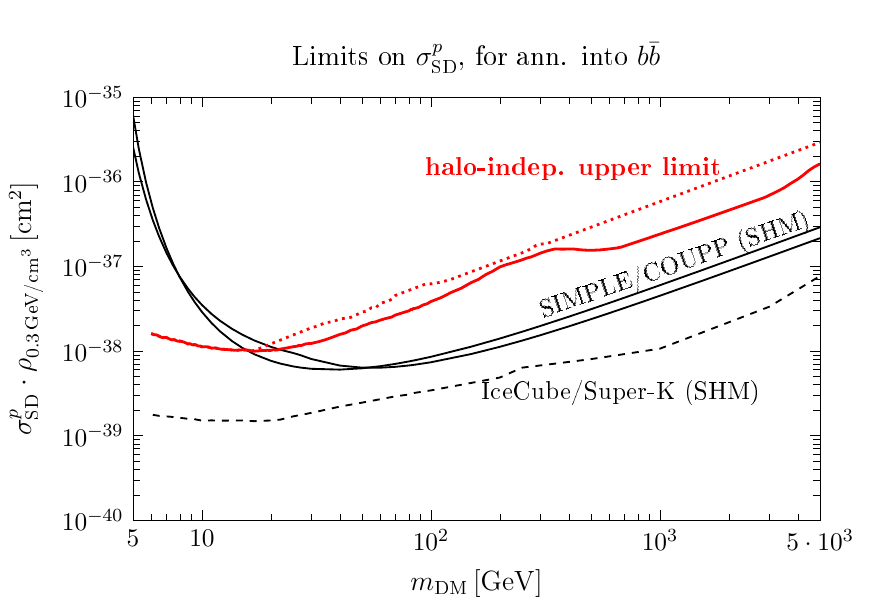}
\end{center}
\caption{ \small Halo-independent upper limits on \sip (upper panel) and \sdp (lower panel), normalized to a local dark matter density $\rho_\text{loc} = 0.3 \text{ GeV}/\text{cm}^3$, for annihilation into $W^+ W^- / \tau^+ \tau^-$ (left panels) and $b \bar{b}$ (right panels). The red solid curves correspond to $v_\text{max} = 777$ km/s, while the red dotted curves assume $v_\text{max} = 0.05 \,$c. Also, the black solid and dashed curves show the upper limits from different direct detection experiments and neutrino telescopes, assuming the Standard Halo Model (SHM) as defined in the text.}
\label{fig:halo_independent_limits}
\end{figure}

The halo-independent upper limits on the spin independent and spin dependent cross sections are shown in the upper and lower plots of Fig.~\ref{fig:halo_independent_limits}, respectively, as a function of the dark matter mass, normalized to a local dark matter density $\rho_\text{loc}= 0.3 \text{ GeV}/\text{cm}^3$. The left panels correspond to dark matter annihilating into $W^+ W^- / \tau^+ \tau^-$, while the right panels assume annihilation into $b \bar{b}$. For each dark matter mass we calculate, using the method described above, an upper limit on the cross section using the null results from one direct detection experiment (either XENON100 or SuperCDMS for the spin-independent scattering, or COUPP or SIMPLE for the spin-dependent scattering\footnote {We do not consider limits from LUX~\cite{Akerib:2013tjd}, as the information provided by the collaboration is not sufficient to derive limits for non-standard velocity distributions~\cite{DelNobile:2013gba}. Similarly, also the recent results of the PICO experiment~\cite{Amole:2015lsj} can not be generalized to other velocity distributions without additional information about efficiency of the detector.}) and one neutrino telescope (either IceCube or SuperKamiokande). Finally, we take the upper limit on the spin-independent and spin-dependent cross sections which is allowed by all the experiments considered in our analysis. 

In order to compare our results with the standard exclusion bounds from direct detection experiments as well as from neutrino telescopes, we also show in Fig.~\ref{fig:halo_independent_limits} the upper limits on the scattering cross section for the specific case of the Standard Halo Model (SHM), i.e. assuming a Maxwell-Boltzmann velocity distribution with velocity dispersion $v_0 = 230$ km/s and galactic escape velocity $v_\text{esc} = 533$ km/s. In the upper panels, corresponding to spin-independent scattering, black solid lines show the limits from Super-CDMS and XENON100, while in the lower panels, corresponding to the case of spin-dependent scattering off protons, they show bounds from SIMPLE and COUPP. Besides, the black dashed lines show the upper limits from Super-K and IceCube. The halo-independent limits are, as expected, somewhat weaker than the combined limit from experiments assuming the standard halo model. Nevertheless, our limits are remarkably strong and reach $\sip\lesssim 10^{-43} \, (10^{-42}) \cm^2$ and  $\sdp\lesssim 10^{-37} \, (3\times 10^{-37 }) \cm^2$, for annihilations into $W^+W^-$ ($b\bar b$) at $m_\DM=1\TeV$, assuming $\rho_\text{loc}=0.3\text{ GeV}/\text{cm}^3$. For the spin independent coupling to protons these limits are better than those obtained by IceCube for the SHM, while for the spin dependent coupling to protons, only a factor of a few worse than the limits from the SIMPLE or COUPP experiments, also assuming the SHM.

The limits presented in this work rely on rather weak assumptions. First, we assume that the dark matter density and velocity distribution at the position of the Sun and the Earth are identical and constant over the equilibration time of captures and annihilations, which we estimate to be $\sim$10-100 million years or about 0.2-2\% of the age of the Sun. We also assume $v_\text{max}=777$ km/s, which follows from the well motivated assumption that all dark matter particles in the Solar System are gravitationally bound to the galaxy. Nevertheless, we have checked that even if the value of the maximal speed is doubled, our limits do not change significantly, since the largest value of the scattering cross-section $\sigma_*$ is typically attained at the velocity $\tilde{v}$, unless $v_\text{max}$ is very large, as can be seen from Fig.~\ref{fig:limits-streams} (we recall that $\tilde{v}$ is defined as the velocity for which $\sigma^\text{DD}_\text{max}(\tilde{v})=\sigma^\text{NT}_\text{max}(\tilde{v})$). To include the possibility of a very large maximal velocity, we have also calculated the corresponding limits for $v_\text{max}=0.05 c$, which approximately corresponds to the largest value of the velocity where the standard non-relativistic description of the scattering and capture rates in Eqs.~(\ref{eq:diff_scattering_rate}), (\ref{eq:general_formula_capture_rate}) can be applied.  The limits are shown in  Fig.~\ref{fig:halo_independent_limits} as a dotted line and are, at most, one order of magnitude worse than those derived for $v_\text{max}=777$ km/s.

The halo-independent upper limits on the scattering cross section presented in this work will improve with upcoming data from direct detection experiments and neutrino telescopes, assuming no signal is detected. It follows from Fig.~\ref{fig:limits-streams} that the limits on the spin-independent cross section can be improved by decreasing the threshold of the corresponding direct detection experiment, or by increasing the sensitivity of neutrino telescopes. In the near future, this could be achievable with e.g. Super-CDMS SNOLAB~\cite{supercdms_snolab} and the results from the 86-string configuration of IceCube~\cite{icecube_86}. On the other hand, for spin-dependent interactions, the current limiting factor is the exposure of the relevant direct detection experiments, as can be seen from the lower panels in Fig.~\ref{fig:limits-streams}. Here, future upgrades of the PICO experiment~\cite{pico_500kg} would help in improving the halo-independent upper limits.

\section{Lower limit on the cross section from signal events}
\label{sec:lower_limit_xenon1t}

We briefly discuss in this section the implications for the determination of the dark matter parameters in the case of a positive signal in a xenon based experiment using our halo-independent approach. In contrast to the previous discussion, in this section we do not consider the complementary information arising from neutrino telescopes. We consider for concreteness an experiment with an exposure of 1 ton$\, \cdot \,$yr, sensitive to the energy range between 3 and 45 keV, with 100\% detection efficiency and an energy resolution $\sigma(E_R) = 0.6 \text{ keV} \sqrt{E_R / \text{keV}}$~\cite{Pato:2010zk}. We also consider a benchmark scenario where the true dark matter parameters are $m_\text{DM}^\text{(true)} = 100$ GeV and $\sigma_\text{SI}^{p\text{(true)}} = 10^{-46} \text{ cm}^2$, which approximately lies one order of magnitude  below the current LUX limit. Assuming the SHM as the true velocity distribution, this benchmark scenario would lead to $\simeq 19$ events in the future experiment, which amounts to  a 90\% C.L. lower limit on the number of recoil events $R_\text{min} = 14.0$. 

\begin{figure}
\begin{center}
\includegraphics[scale=1.3]{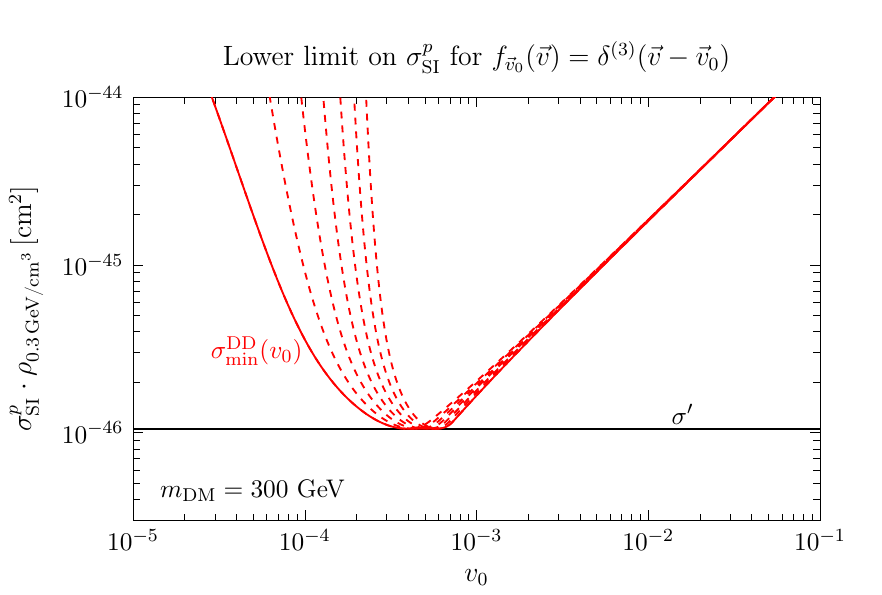}
\end{center}
\caption{\small Lower limit on $\sip$, normalized to a local dark matter density $\rho_\text{loc} = 0.3 \text{ GeV}/\text{cm}^3$, derived from the hypothetical detection of 19 recoil events at a xenon based direct detection experiment, for $m_\text{DM}=300$ GeV and for a velocity distribution corresponding to a stream with speed $v_0$ with respect to the rest frame of the Sun, $f_{\vec v_0}(v)=\delta^{(3)}(\vec v-\vec v_0)$. The minimum value of the cross section leading to the hypothetical number of recoil events is denoted by $\sigma'$.}
\label{fig:XENON1T-signal}
\end{figure}

On the basis of this number of events, we now calculate a halo-independent \emph{lower} limit on the scattering cross section following a similar rationale as in Section \ref{sec:superposition}. We first determine, for a given dark matter mass, a lower limit on the scattering cross section assuming a dark matter stream with velocity $\vec{v}_0$ with respect to the Sun from requiring $R_{\vec{v}_0} \geq R_\text{min}$, and we denote $\sigma^\text{DD}_\text{min}(v_0)$ the strongest among the limits for all possible angles between the stream velocity $\vec v_0$ and a fixed Earth velocity $\vec v_E$. Then, by construction, $R_{\vec{v}_0} (\sigma) \leq R_\text{min}$ for $\sigma \leq \sigma_\text{min}^{\text{DD}}(v_0)$. This procedure is illustrated in Fig.~\ref{fig:XENON1T-signal} for $m_\text{DM} = 300$ GeV. The dashed red lines show the limits for various angles between the stream velocity $\vec v_0$ and a fixed Earth velocity $\vec v_E$, while the red solid line corresponds to $\sigma^\text{DD}_\text{min}(v_0)$. Finally we construct, following similar steps as in Section \ref{sec:superposition}, a lower limit on the scattering cross section for a general velocity distribution from the corresponding limits for the stream-like distributions:
\begin{align}
\sigma \geq \left[\int_{|\vec v_0|\leq v_\text{max}}  \text{d}^3 v_0 \frac{f(\vec v_0)}{\sigma^{\rm DD}_\text{min}(v_0)}\right]^{-1}\;.
\label{eq:sigmaDD_general_lower_limit}
\end{align}
For each dark matter mass the function $\sigma_\text{min}^{\text{DD}}(v_0)$ has a minimum, which we denote as $\sigma'$ (see Fig.~\ref{fig:XENON1T-signal}). Then, from Eqs.~(\ref{eq:normalization}) and (\ref{eq:sigmaDD_general_lower_limit}) we obtain the following lower limit on the scattering cross section, which is independent of the velocity distribution:
\begin{align}
\sigma\geq \sigma'\;.
\end{align}

\begin{figure}
\begin{center}
\includegraphics[scale=1.3]{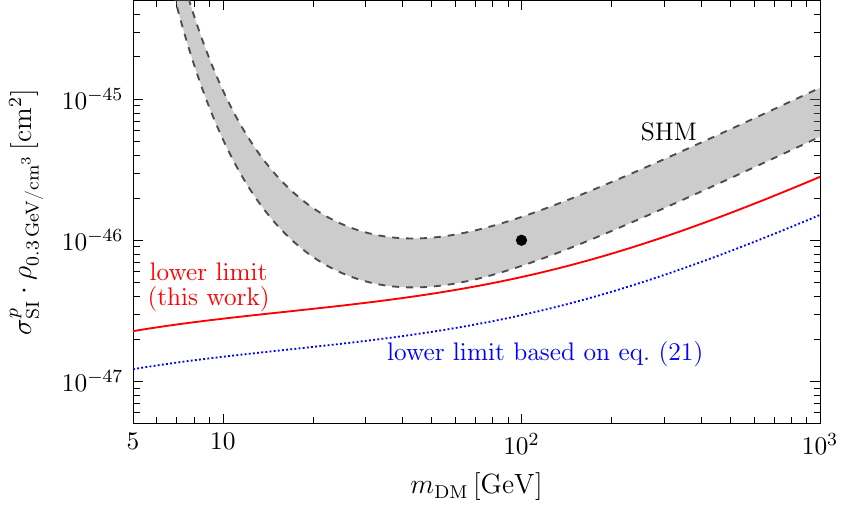}
\end{center}
\caption{ \small Halo-independent lower limit on $\sip$, normalized to a local dark matter density $\rho_\text{loc} = 0.3 \text{ GeV}/\text{cm}^3$, from the hypothetical observation of 19 recoil events in a xenon based direct detection experiment (red line), reconstructed 90\% C.L. region assuming the SHM (gray band), and halo-independent lower limit derived from Eq.~(\ref{eq:limit_Blennow_et_al}) (blue dotted line). The black dot indicates the true dark matter mass and scattering cross section.}
\label{fig:x-section_lower_limit}
\end{figure}

Fig.~\ref{fig:x-section_lower_limit} shows, as a red solid line, the lower limit on the spin independent scattering cross section with protons as a function of the dark matter mass, calculated using this method. For comparison, the Figure also shows, as a gray band, the reconstructed 90\% C.L. region assuming the SHM, as well as the halo-independent lower bound recently proposed in \cite{Blennow:2015gta} based on the general inequality
\begin{align}
\int_{v > v_\text{min}} \text{d}^3 v \, \frac{f (\vec{v})}{v} \leq \frac{1}{v_\text{min}} \,,
\label{eq:limit_Blennow_et_al}
\end{align}
where $v_\text{min}$ is the minimal speed necessary for a dark matter particle to induce a recoil with energy above the threshold of the experiment ({\it cf.} Section \ref{sec:formalism}). Notably, for the true dark matter mass, the lower limit lies only a factor of $\sim 2$ below the true cross section (shown as a black point). We have checked that this conclusion does not change significantly with the number of observed events, as long as it is larger than $\simeq 5$. Furthermore, had the true dark matter mass been different, for the same number of signal events one finds a lower limit on the cross section which lies only a factor $\sim 2-3$ below the true value, provided $m_\text{DM}^{\text{(true)}}\gtrsim 30$ GeV. This is due to the degeneracy between the dark matter mass and the scattering cross section in the event rate, which is reflected for the SHM in the gray band in Fig.~\ref{fig:x-section_lower_limit}.

Finally, we note that our method only uses the information about the total number of events, and hence it is particularly well-suited when only a few events are detected. On the other hand, once enough data is available to infer the recoil energy spectrum, halo-independent methods using the spectral information can break the degeneracy between the dark matter mass and scattering cross section~\cite{Kavanagh:2013wba,Feldstein:2014gza}.

\section{Conclusions}
\label{sec:conclusions}

We have presented a new method to calculate limits on the dark matter scattering cross section for a general dark matter velocity distribution. Our method is based on decomposing the velocity
distribution in dark matter streams with fixed velocity. We have then calculated the upper limit on the scattering cross section as a function of the stream speed for various direct detection experiments and neutrino telescopes and we have presented a simple formula that allows to calculate the corresponding limit for a general velocity distribution, based on this decomposition. This formula can be useful for a fast calculation of the limits for a general velocity distribution, since all the details of the calculation of the number of recoil events for a given direct detection experiment or of the capture rate in the Sun are already included in the limits for the stream distributions. 

We have applied this method to calculate an upper limit on the dark matter scattering cross section by combining the null results from direct detection experiments and from neutrino telescopes and which is independent of the dark matter velocity distribution. Our only assumptions are that the dark matter density and velocity distribution at the position of the Sun and the Earth are identical and constant over the last 10-100 million years, and that the dark matter particles move non-relativistically. The resulting limits are remarkably strong and reach $\sip\lesssim 10^{-43} \, (10^{-42}) \cm^2$ and  $\sdp\lesssim 10^{-37} \, (3\times 10^{-37 }) \cm^2$, for annihilations into $W^+W^-$ ($b\bar b$) at $m_\DM=1\TeV$, assuming $\rho_\text{loc}=0.3 \text{ GeV}/\text{cm}^3$. 

Lastly, we have presented a simple procedure to derive a halo-independent lower limit on the scattering cross section as a function of the dark matter mass in the case of the detection of signal events in a future experiment. We have illustrated this procedure for a ton-scale xenon based experiment and we have found a lower limit which lies, for a true dark matter mass larger than $\sim 30$ GeV, only a factor of a few below the true cross section. 

\section*{Acknowledgements}

FF is grateful to TUM for hospitality. The work of FF has been partially 
supported by the DOE at WU, and the work of AI and SW has been partially supported by the
DFG cluster of excellence “Origin and Structure of the Universe,” the TUM Graduate School and
the Studienstiftung des Deutschen Volkes.

\appendix
\section{Calculating upper limits from direct detection data}
\label{sec:appendix_dd}

In this appendix we provide the details of our method of deriving upper limits on the scattering cross section from XENON100, SuperCDMS, COUPP, and SIMPLE. For each of these experiments, we obtain the upper limit by comparing the number of expected recoil events $R$, given by Eq.~(\ref{eq:diff_scattering_rate}), with the maximal number of events $R_\text{max}$ allowed by the result of the experiment.

For XENON100, we calculate the detection efficiency $\epsilon (E_R)$ following~\cite{Aprile:2011hx}. More precisely, we use the scintillation efficiency $\mathcal{L}_\text{eff} (E_R)$ from~\cite{Aprile:2011hi}, conservatively setting it to zero below $E_R = 3$ keV. The detection efficiency as a function of the S1 signal, as well as the efficiency related to the S2 threshold cut is taken from~\cite{Aprile:2012nq}. After running with an exposure of 7636.4 kg$\,\cdot\,$days, XENON100 observed two events in the predefined signal region given by $S_1^{\text{min}} = 3$ PE and $S_1^{\text{max}} = 30$ PE. Making no assumption about the background, the 90\% C.L. upper limit on the number of recoil events is given by $R_\text{max} = 5.32$.

The SuperCDMS limits are obtained by using the efficiency $\epsilon (E_R)$ given in~\cite{Agnese:2014aze} in the recoil energy range between 1.6 keV and 10 keV. With an exposure of 577 kg$\,\cdot\,$days, SuperCDMS observed 11 candidate events. As there is no full understanding of the expected background, we derive limits by conservatively assuming  zero background events, leading to $R_\text{max} = 16.6$ at 90\% C.L.

The latest result of the COUPP collaboration~\cite{Behnke:2012ys} is presented in the form of three non-overlapping data sets with different nucleation thresholds, given by 7.8 keV, 11.0 keV, and 15.5 keV. The effective exposures (number of observed events) are 55.8 kg$\,\cdot\,$days (2 events), 70.0 kg$\,\cdot\,$days (3 events), and 311.7 kg$\,\cdot\,$days (8 events), respectively. Making no assumption about the background, the maximal number of expected events at 90\% C.L. is given by $R_\text{max} = 5.32, \, 6.68,$ and 12.99, respectively. We derive individual upper limits for each of the three data sets, choosing the most constraining one separately for each dark matter mass and interaction type. In all cases, the efficiency is assumed to be 0.49 above threshold for scattering off C or F, and 1 for scattering off I~\cite{Behnke:2012ys}; furthermore, we only consider recoils up to 500 keV.

Finally, we derive limits from the SIMPLE experiment by using the combined Stage I and II results presented in~\cite{Felizardo:2011uw}. Following the method pursued by the collaboration, we use an energy threshold of $E_T = 8$ keV, with an efficiency $\epsilon( E_R ) = 1-\exp\left(- 3.6 \left( E_R / E_T -1 \right) \right)$. As for COUPP, we limit ourselves to recoil energies below 500 keV. After running with an effective exposure of 20.18 kg$\,\cdot\,$days, 11 events were observed. With the same conservative assumption $b = 14.53$ for the expected number of background events as used in~\cite{Felizardo:2011uw}, one obtains $R_\text{max} = 4.02$ at 90\% C.L, using the Feldman-Cousins procedure~\cite{Feldman:1997qc}.

For all experiments considered in this work, we checked that, under the assumptions specified above, the upper limit derived using the standard Maxwell-Boltzmann distribution matches sufficiently well the limit published by the corresponding collaboration.

\bibliographystyle{JHEP-mod}
\bibliography{references}

\providecommand{\href}[2]{#2}\begingroup\raggedright\begin{thebibliography}{10}

\bibitem{Ade:2015xua}
{\bf Planck}, P.~Ade {\em et.~al.}, {\it {Planck 2015 results. XIII.
  Cosmological parameters}},  \href{http://xxx.lanl.gov/abs/1502.01589}{{\tt
  arXiv:1502.01589}}.

\bibitem{Bertone:2010zza}
G.~Bertone, J.~Silk, B.~Moore, J.~Diemand, J.~Bullock, {\em et.~al.}, {\it
  {Particle Dark Matter: Observations, Models and Searches}}, .

\bibitem{Bergstrom:2000pn}
L.~Bergstrom, {\it {Nonbaryonic dark matter: Observational evidence and
  detection methods}},  {\em Rept.Prog.Phys.} {\bf 63} (2000) 793,
  [\href{http://xxx.lanl.gov/abs/hep-ph/0002126}{{\tt hep-ph/0002126}}].

\bibitem{Jungman:1995df}
G.~Jungman, M.~Kamionkowski, and K.~Griest, {\it {Supersymmetric dark matter}},
   {\em Phys.Rept.} {\bf 267} (1996) 195--373,
  [\href{http://xxx.lanl.gov/abs/hep-ph/9506380}{{\tt hep-ph/9506380}}].

\bibitem{Bertone:2004pz}
G.~Bertone, D.~Hooper, and J.~Silk, {\it {Particle dark matter: Evidence,
  candidates and constraints}},  {\em Phys.Rept.} {\bf 405} (2005) 279--390,
  [\href{http://xxx.lanl.gov/abs/hep-ph/0404175}{{\tt hep-ph/0404175}}].

\bibitem{Goodman:1984dc}
M.~W. Goodman and E.~Witten, {\it {Detectability of Certain Dark Matter
  Candidates}},  {\em Phys.Rev.} {\bf D31} (1985) 3059.

\bibitem{Silk:1985ax}
J.~Silk, K.~A. Olive, and M.~Srednicki, {\it {The Photino, the Sun and
  High-Energy Neutrinos}},  {\em Phys.Rev.Lett.} {\bf 55} (1985) 257--259.

\bibitem{Akerib:2013tjd}
{\bf LUX Collaboration}, D.~Akerib {\em et.~al.}, {\it {First results from the
  LUX dark matter experiment at the Sanford Underground Research Facility}},
  {\em Phys.Rev.Lett.} {\bf 112} (2014), no.~9 091303,
  [\href{http://xxx.lanl.gov/abs/1310.8214}{{\tt arXiv:1310.8214}}].

\bibitem{Aprile:2012nq}
{\bf XENON100}, E.~Aprile {\em et.~al.}, {\it {Dark Matter Results from 225
  Live Days of XENON100 Data}},  {\em Phys.Rev.Lett.} {\bf 109} (2012) 181301,
  [\href{http://xxx.lanl.gov/abs/1207.5988}{{\tt arXiv:1207.5988}}].

\bibitem{Agnese:2014aze}
{\bf SuperCDMS}, R.~Agnese {\em et.~al.}, {\it {Search for Low-Mass Weakly
  Interacting Massive Particles with SuperCDMS}},  {\em Phys.Rev.Lett.} {\bf
  112} (2014), no.~24 241302, [\href{http://xxx.lanl.gov/abs/1402.7137}{{\tt
  arXiv:1402.7137}}].

\bibitem{Amole:2015lsj}
{\bf PICO}, C.~Amole {\em et.~al.}, {\it {Dark Matter Search Results from the
  PICO-2L C$_3$F$_8$ Bubble Chamber}},
  \href{http://xxx.lanl.gov/abs/1503.00008}{{\tt arXiv:1503.00008}}.

\bibitem{Behnke:2012ys}
{\bf COUPP}, E.~Behnke {\em et.~al.}, {\it {First Dark Matter Search Results
  from a 4-kg CF$_3$I Bubble Chamber Operated in a Deep Underground Site}},
  {\em Phys.Rev.} {\bf D86} (2012), no.~5 052001,
  [\href{http://xxx.lanl.gov/abs/1204.3094}{{\tt arXiv:1204.3094}}].

\bibitem{Felizardo:2011uw}
M.~Felizardo, T.~Girard, T.~Morlat, A.~Fernandes, A.~Ramos, {\em et.~al.}, {\it
  {Final Analysis and Results of the Phase II SIMPLE Dark Matter Search}},
  {\em Phys.Rev.Lett.} {\bf 108} (2012) 201302,
  [\href{http://xxx.lanl.gov/abs/1106.3014}{{\tt arXiv:1106.3014}}].

\bibitem{Choi:2015ara}
{\bf Super-Kamiokande}, K.~Choi {\em et.~al.}, {\it {Search for neutrinos from
  annihilation of captured low-mass dark matter particles in the Sun by
  Super-Kamiokande}},  {\em Phys.Rev.Lett.} {\bf 114} (2015), no.~14 141301,
  [\href{http://xxx.lanl.gov/abs/1503.04858}{{\tt arXiv:1503.04858}}].

\bibitem{Aartsen:2012kia}
{\bf IceCube}, M.~Aartsen {\em et.~al.}, {\it {Search for dark matter
  annihilations in the Sun with the 79-string IceCube detector}},  {\em
  Phys.Rev.Lett.} {\bf 110} (2013), no.~13 131302,
  [\href{http://xxx.lanl.gov/abs/1212.4097}{{\tt arXiv:1212.4097}}].

\bibitem{Drukier:1986tm}
A.~Drukier, K.~Freese, and D.~Spergel, {\it {Detecting Cold Dark Matter
  Candidates}},  {\em Phys.Rev.} {\bf D33} (1986) 3495--3508.

\bibitem{Kuhlen:2009vh}
M.~Kuhlen, N.~Weiner, J.~Diemand, P.~Madau, B.~Moore, {\em et.~al.}, {\it {Dark
  Matter Direct Detection with Non-Maxwellian Velocity Structure}},  {\em JCAP}
  {\bf 1002} (2010) 030, [\href{http://xxx.lanl.gov/abs/0912.2358}{{\tt
  arXiv:0912.2358}}].

\bibitem{Lisanti:2010qx}
M.~Lisanti, L.~E. Strigari, J.~G. Wacker, and R.~H. Wechsler, {\it {The Dark
  Matter at the End of the Galaxy}},  {\em Phys.Rev.} {\bf D83} (2011) 023519,
  [\href{http://xxx.lanl.gov/abs/1010.4300}{{\tt arXiv:1010.4300}}].

\bibitem{Mao:2012hf}
Y.-Y. Mao, L.~E. Strigari, R.~H. Wechsler, H.-Y. Wu, and O.~Hahn, {\it
  {Halo-to-Halo Similarity and Scatter in the Velocity Distribution of Dark
  Matter}},  {\em Astrophys.J.} {\bf 764} (2013) 35,
  [\href{http://xxx.lanl.gov/abs/1210.2721}{{\tt arXiv:1210.2721}}].

\bibitem{Read:2008fh}
J.~Read, G.~Lake, O.~Agertz, and V.~P. Debattista, {\it {Thin, thick and dark
  discs in LCDM}},  {\em Mon.Not.Roy.Astron.Soc.} {\bf 389} (2008) 1041--1057,
  [\href{http://xxx.lanl.gov/abs/0803.2714}{{\tt arXiv:0803.2714}}].

\bibitem{Read:2009iv}
J.~Read, L.~Mayer, A.~Brooks, F.~Governato, and G.~Lake, {\it {A dark matter
  disc in three cosmological simulations of Milky Way mass galaxies}},  {\em
  Mon.Not.Roy.Astron.Soc.} {\bf 397} (2009) 44,
  [\href{http://xxx.lanl.gov/abs/0902.0009}{{\tt arXiv:0902.0009}}].

\bibitem{Purcell:2009yp}
C.~W. Purcell, J.~S. Bullock, and M.~Kaplinghat, {\it {The Dark Disk of the
  Milky Way}},  {\em Astrophys.J.} {\bf 703} (2009) 2275--2284,
  [\href{http://xxx.lanl.gov/abs/0906.5348}{{\tt arXiv:0906.5348}}].

\bibitem{Ling:2009eh}
F.~Ling, E.~Nezri, E.~Athanassoula, and R.~Teyssier, {\it {Dark Matter Direct
  Detection Signals inferred from a Cosmological N-body Simulation with
  Baryons}},  {\em JCAP} {\bf 1002} (2010) 012,
  [\href{http://xxx.lanl.gov/abs/0909.2028}{{\tt arXiv:0909.2028}}].

\bibitem{Fox:2010bu}
P.~J. Fox, G.~D. Kribs, and T.~M. Tait, {\it {Interpreting Dark Matter Direct
  Detection Independently of the Local Velocity and Density Distribution}},
  {\em Phys.Rev.} {\bf D83} (2011) 034007,
  [\href{http://xxx.lanl.gov/abs/1011.1910}{{\tt arXiv:1011.1910}}].

\bibitem{Fox:2010bz}
P.~J. Fox, J.~Liu, and N.~Weiner, {\it {Integrating Out Astrophysical
  Uncertainties}},  {\em Phys.Rev.} {\bf D83} (2011) 103514,
  [\href{http://xxx.lanl.gov/abs/1011.1915}{{\tt arXiv:1011.1915}}].

\bibitem{McCabe:2010zh}
C.~McCabe, {\it {The Astrophysical Uncertainties Of Dark Matter Direct
  Detection Experiments}},  {\em Phys.Rev.} {\bf D82} (2010) 023530,
  [\href{http://xxx.lanl.gov/abs/1005.0579}{{\tt arXiv:1005.0579}}].

\bibitem{McCabe:2011sr}
C.~McCabe, {\it {DAMA and CoGeNT without astrophysical uncertainties}},  {\em
  Phys.Rev.} {\bf D84} (2011) 043525,
  [\href{http://xxx.lanl.gov/abs/1107.0741}{{\tt arXiv:1107.0741}}].

\bibitem{Frandsen:2011gi}
M.~T. Frandsen, F.~Kahlhoefer, C.~McCabe, S.~Sarkar, and K.~Schmidt-Hoberg,
  {\it {Resolving astrophysical uncertainties in dark matter direct
  detection}},  {\em JCAP} {\bf 1201} (2012) 024,
  [\href{http://xxx.lanl.gov/abs/1111.0292}{{\tt arXiv:1111.0292}}].

\bibitem{Gondolo:2012rs}
P.~Gondolo and G.~B. Gelmini, {\it {Halo independent comparison of direct dark
  matter detection data}},  {\em JCAP} {\bf 1212} (2012) 015,
  [\href{http://xxx.lanl.gov/abs/1202.6359}{{\tt arXiv:1202.6359}}].

\bibitem{HerreroGarcia:2012fu}
J.~Herrero-Garcia, T.~Schwetz, and J.~Zupan, {\it {Astrophysics independent
  bounds on the annual modulation of dark matter signals}},  {\em
  Phys.Rev.Lett.} {\bf 109} (2012) 141301,
  [\href{http://xxx.lanl.gov/abs/1205.0134}{{\tt arXiv:1205.0134}}].

\bibitem{DelNobile:2013cva}
E.~Del~Nobile, G.~Gelmini, P.~Gondolo, and J.-H. Huh, {\it {Generalized Halo
  Independent Comparison of Direct Dark Matter Detection Data}},  {\em JCAP}
  {\bf 1310} (2013) 048, [\href{http://xxx.lanl.gov/abs/1306.5273}{{\tt
  arXiv:1306.5273}}].

\bibitem{Fox:2014kua}
P.~J. Fox, Y.~Kahn, and M.~McCullough, {\it {Taking Halo-Independent Dark
  Matter Methods Out of the Bin}},  {\em JCAP} {\bf 1410} (2014), no.~10 076,
  [\href{http://xxx.lanl.gov/abs/1403.6830}{{\tt arXiv:1403.6830}}].

\bibitem{Feldstein:2014gza}
B.~Feldstein and F.~Kahlhoefer, {\it {A new halo-independent approach to dark
  matter direct detection analysis}},  {\em JCAP} {\bf 1408} (2014) 065,
  [\href{http://xxx.lanl.gov/abs/1403.4606}{{\tt arXiv:1403.4606}}].

\bibitem{Feldstein:2014ufa}
B.~Feldstein and F.~Kahlhoefer, {\it {Quantifying (dis)agreement between direct
  detection experiments in a halo-independent way}},  {\em JCAP} {\bf 1412}
  (2014), no.~12 052, [\href{http://xxx.lanl.gov/abs/1409.5446}{{\tt
  arXiv:1409.5446}}].

\bibitem{Anderson:2015xaa}
A.~J. Anderson, P.~J. Fox, Y.~Kahn, and M.~McCullough, {\it {Halo-Independent
  Direct Detection Analyses Without Mass Assumptions}},
  \href{http://xxx.lanl.gov/abs/1504.03333}{{\tt arXiv:1504.03333}}.

\bibitem{Bozorgnia:2014gsa}
N.~Bozorgnia and T.~Schwetz, {\it {What is the probability that direct
  detection experiments have observed Dark Matter?}},  {\em JCAP} {\bf 1412}
  (2014), no.~12 015, [\href{http://xxx.lanl.gov/abs/1410.6160}{{\tt
  arXiv:1410.6160}}].

\bibitem{Kavanagh:2014rya}
B.~J. Kavanagh, M.~Fornasa, and A.~M. Green, {\it {Probing WIMP particle
  physics and astrophysics with direct detection and neutrino telescope data}},
   {\em Phys.Rev.} {\bf D91} (2015), no.~10 103533,
  [\href{http://xxx.lanl.gov/abs/1410.8051}{{\tt arXiv:1410.8051}}].

\bibitem{Blennow:2015oea}
M.~Blennow, J.~Herrero-Garcia, and T.~Schwetz, {\it {A halo-independent lower
  bound on the dark matter capture rate in the Sun from a direct detection
  signal}},  {\em JCAP} {\bf 1505} (2015), no.~05 036,
  [\href{http://xxx.lanl.gov/abs/1502.03342}{{\tt arXiv:1502.03342}}].

\bibitem{Piffl:2013mla}
T.~Piffl, C.~Scannapieco, J.~Binney, M.~Steinmetz, R.-D. Scholz, {\em et.~al.},
  {\it {The RAVE survey: the Galactic escape speed and the mass of the Milky
  Way}},  {\em Astron.Astrophys.} {\bf 562} (2014) A91,
  [\href{http://xxx.lanl.gov/abs/1309.4293}{{\tt arXiv:1309.4293}}].

\bibitem{Xue:2008se}
{\bf SDSS}, X.~Xue {\em et.~al.}, {\it {The Milky Way's Circular Velocity Curve
  to 60 kpc and an Estimate of the Dark Matter Halo Mass from Kinematics of
  ~2400 SDSS Blue Horizontal Branch Stars}},  {\em Astrophys.J.} {\bf 684}
  (2008) 1143--1158, [\href{http://xxx.lanl.gov/abs/0801.1232}{{\tt
  arXiv:0801.1232}}].

\bibitem{McMillan:2009yr}
P.~J. McMillan and J.~J. Binney, {\it {The uncertainty in Galactic
  parameters}},  {\em Mon.Not.Roy.Astron.Soc.} {\bf 402} (2010) 934,
  [\href{http://xxx.lanl.gov/abs/0907.4685}{{\tt arXiv:0907.4685}}].

\bibitem{Bovy:2009dr}
J.~Bovy, D.~W. Hogg, and H.-W. Rix, {\it {Galactic masers and the Milky Way
  circular velocity}},  {\em Astrophys.J.} {\bf 704} (2009) 1704--1709,
  [\href{http://xxx.lanl.gov/abs/0907.5423}{{\tt arXiv:0907.5423}}].

\bibitem{Gould:1987ir}
A.~Gould, {\it {Resonant Enhancements in WIMP Capture by the Earth}},  {\em
  Astrophys.J.} {\bf 321} (1987) 571.

\bibitem{Cerdeno:2010jj}
D.~G. Cerdeno and A.~M. Green, {\it {Direct detection of WIMPs}},
  \href{http://xxx.lanl.gov/abs/1002.1912}{{\tt arXiv:1002.1912}}.

\bibitem{Lewin:1995rx}
J.~Lewin and P.~Smith, {\it {Review of mathematics, numerical factors, and
  corrections for dark matter experiments based on elastic nuclear recoil}},
  {\em Astropart.Phys.} {\bf 6} (1996) 87--112.

\bibitem{Bednyakov:2006ux}
V.~Bednyakov and F.~Simkovic, {\it {Nuclear spin structure in dark matter
  search: The Finite momentum transfer limit}},  {\em Phys.Part.Nucl.} {\bf 37}
  (2006) S106--S128, [\href{http://xxx.lanl.gov/abs/hep-ph/0608097}{{\tt
  hep-ph/0608097}}].

\bibitem{Gondolo:2004sc}
P.~Gondolo, J.~Edsjo, P.~Ullio, L.~Bergstrom, M.~Schelke, {\em et.~al.}, {\it
  {DarkSUSY: Computing supersymmetric dark matter properties numerically}},
  {\em JCAP} {\bf 0407} (2004) 008,
  [\href{http://xxx.lanl.gov/abs/astro-ph/0406204}{{\tt astro-ph/0406204}}].

\bibitem{Catena:2015uha}
R.~Catena and B.~Schwabe, {\it {Form factors for dark matter capture by the Sun
  in effective theories}},  \href{http://xxx.lanl.gov/abs/1501.03729}{{\tt
  arXiv:1501.03729}}.

\bibitem{Serenelli:2009yc}
A.~Serenelli, S.~Basu, J.~W. Ferguson, and M.~Asplund, {\it {New Solar
  Composition: The Problem With Solar Models Revisited}},  {\em Astrophys.J.}
  {\bf 705} (2009) L123--L127, [\href{http://xxx.lanl.gov/abs/0909.2668}{{\tt
  arXiv:0909.2668}}].

\bibitem{DelNobile:2013gba}
E.~Del~Nobile, G.~B. Gelmini, P.~Gondolo, and J.-H. Huh, {\it {Update on Light
  WIMP Limits: LUX, lite and Light}},  {\em JCAP} {\bf 1403} (2014) 014,
  [\href{http://xxx.lanl.gov/abs/1311.4247}{{\tt arXiv:1311.4247}}].

\bibitem{supercdms_snolab}
http://cdms.berkeley.edu/scdmssnolab.html.
\newblock Accessed: 2015-08-18.

\bibitem{icecube_86}
WIMP Annihilations in the Sun : A search using first year of operation of the
  completed IceCube neutrino telescope.
\newblock Talk by M. Rameez at Moriond 2015.

\bibitem{pico_500kg}
http://www.picoexperiment.com/.
\newblock Accessed: 2015-08-18.

\bibitem{Pato:2010zk}
M.~Pato, L.~Baudis, G.~Bertone, R.~Ruiz~de Austri, L.~E. Strigari, {\em
  et.~al.}, {\it {Complementarity of Dark Matter Direct Detection Targets}},
  {\em Phys.Rev.} {\bf D83} (2011) 083505,
  [\href{http://xxx.lanl.gov/abs/1012.3458}{{\tt arXiv:1012.3458}}].

\bibitem{Blennow:2015gta}
M.~Blennow, J.~Herrero-Garcia, T.~Schwetz, and S.~Vogl, {\it {Halo-independent
  tests of dark matter direct detection signals: local DM density, LHC, and
  thermal freeze-out}},  \href{http://xxx.lanl.gov/abs/1505.05710}{{\tt
  arXiv:1505.05710}}.

\bibitem{Kavanagh:2013wba}
B.~J. Kavanagh and A.~M. Green, {\it {Model independent determination of the
  dark matter mass from direct detection experiments}},  {\em Phys.Rev.Lett.}
  {\bf 111} (2013), no.~3 031302,
  [\href{http://xxx.lanl.gov/abs/1303.6868}{{\tt arXiv:1303.6868}}].

\bibitem{Aprile:2011hx}
{\bf XENON100}, E.~Aprile {\em et.~al.}, {\it {Likelihood Approach to the First
  Dark Matter Results from XENON100}},  {\em Phys.Rev.} {\bf D84} (2011)
  052003, [\href{http://xxx.lanl.gov/abs/1103.0303}{{\tt arXiv:1103.0303}}].

\bibitem{Aprile:2011hi}
{\bf XENON100}, E.~Aprile {\em et.~al.}, {\it {Dark Matter Results from 100
  Live Days of XENON100 Data}},  {\em Phys.Rev.Lett.} {\bf 107} (2011) 131302,
  [\href{http://xxx.lanl.gov/abs/1104.2549}{{\tt arXiv:1104.2549}}].

\bibitem{Feldman:1997qc}
G.~J. Feldman and R.~D. Cousins, {\it {A Unified approach to the classical
  statistical analysis of small signals}},  {\em Phys.Rev.} {\bf D57} (1998)
  3873--3889, [\href{http://xxx.lanl.gov/abs/physics/9711021}{{\tt
  physics/9711021}}].

\end{thebibliography}\endgroup

\end{document}